\documentclass[12pt]{iopart}

\usepackage{amssymb}
\usepackage{iopams}
\usepackage{setstack}
\usepackage{graphicx}

\begin{document}

\title{A model for signal transduction during quorum sensing in \emph{Vibrio harveyi}}

\author{Suman K Banik\footnote{SKB and ATF contributed equally to this work}
\footnote{Present Address: Department of Chemistry, Bose Institute, Kolkata 700 009, India.},
Andrew T Fenley$\ddagger$ and
Rahul V Kulkarni\footnote{Corresponding author}
}

\address{Department of Physics,
Virginia Polytechnic Institute and State University,
Blacksburg, VA 24061-0435, USA.}

\ead{skbanik@vt.edu, afenley@vt.edu, kulkarni@vt.edu}

\begin{abstract}
We present a framework for analyzing luminescence
regulation during
quorum sensing in the bioluminescent bacterium \emph{Vibrio harveyi}.
Using a simplified model for signal transduction in the quorum sensing
pathway, we identify key dimensionless parameters that control the
system's response. These parameters are estimated using experimental
data on luminescence phenotypes for different mutant strains.  The
corresponding model predictions are consistent with results from other
experiments which did not serve as inputs for determining model
parameters. Furthermore, the proposed framework leads to novel
testable predictions for luminescence phenotypes and for responses of
the network to different perturbations.
\end{abstract}




\section{Introduction}

Bacterial survival critically depends on regulatory networks which
integrate multiple inputs to implement important cellular decisions.
A prominent example is the global regulatory network involved in
``quorum sensing'', commonly defined as the regulation of gene
expression in response to cell density.  During the process of quorum
sensing (QS), bacteria produce, secrete and detect signaling molecules
called autoinducers (Miller and Bassler 2001; Waters and Bassler 2005;
Bassler and Losick 2006).  These signals are then processed by the QS
pathway to regulate critical bacterial processes such as biofilm
formation and virulence. The observation that quorum sensing is linked
to both biofilm formation and virulence factor production suggests
that many virulent bacteria can be rendered nonpathogenic by the
inhibition of their QS pathways (Bjarnsholt and Givskov 2007).
Quantitative modeling of the QS pathway can thus provide useful inputs for
treating many common and damaging bacterial infections.

One of the most studied model organisms for QS based regulation is the
bioluminescent marine bacterium \emph{Vibrio harveyi} (Nealson et al., 1970). 
Experimental
studies have led to a detailed characterization of regulatory elements
in the pathway (Henke and Bassler 2004; Mok et al., 2003; Timmen et
al., 2006; Waters and Bassler 2006; Tu and Bassler 2007). The network
(see Fig.~1) includes multiple autoinducers and corresponding sensor
proteins which act together to control the phosphorylation of the
response regulator protein LuxO.  The phosphorylated form of LuxO
(LuxO-P) activates the production of multiple small RNA (sRNA)s which in turn
post-transcriptionally repress the QS master regulatory protein LuxR.
At low cell density, the sRNAs are activated and act to
effectively repress LuxR expression. In contrast, sRNA
production is significantly reduced at high cell density, thereby
giving rise to increased levels of LuxR which leads to the
activation of luminescence genes. The corresponding luminescence
output per cell
profile (i.e., colony luminescence/cell output as a function of cell density)
is frequently used as a reporter to characterize the state of the QS
pathway.

Recent experiments (Henke and Bassler 2004) have analyzed the effects
of mutagenesis of different pathway components on the corresponding
luminescence profile in \emph{V. harveyi}. It was observed that
there are distinct luminescence profiles as the network is perturbed
corresponding to different pathway mutants. The changes in the
luminescence profile were used to infer pathway characteristics such
as relative kinase strengths for the different sensors. Given the
complexity of the network which involves integration of multiple
inputs, it would be desirable to develop a quantitative framework for
inferring pathway characteristics based on network perturbations. The
corresponding quantitative model can then be used to make testable
predictions for future experiments as well as to further analyze existing
experimental data. The aim of this work is to develop such a minimal model
for the QS pathway in \emph{V. harveyi}.

The starting point of our analysis is the observation that
luminescence/cell output is controlled by the degree of phosphorylation of
the response regulator LuxO. We thus develop a simplified model which
connects external autoinducer concentrations to the degree of
phosphorylation of LuxO for the wild type (WT) strain and for
different mutants.  Our analysis identifies key dimensionless
parameters which control the system response and which can be
determined using the experimental results for luminescence
phenotypes. Determination of the effective parameters, in turn, leads
to predictions for the systems response to a broader range of
perturbations, i.e., perturbations distinct from those used to infer
the effective parameters. The corresponding analysis sheds light on
previously obtained experimental results and also gives rise to
testable predictions for future experiments.

The rest of the paper is organized as follows. In Section 2 we give an
overview of the QS network in \emph{V. harveyi}. We then develop a minimal
model of the QS pathway and define key dimensionless parameters which
control the network response characteristics. In Section 3, we connect our
model to experimental data on different luminescence curves and thereby
determine model parameters. In Section 4, we discuss experimentally
testable predictions based on the model and conclude with a summary.


\begin{figure}
\label{fig1}
\includegraphics[width=1.0\linewidth,angle=0]{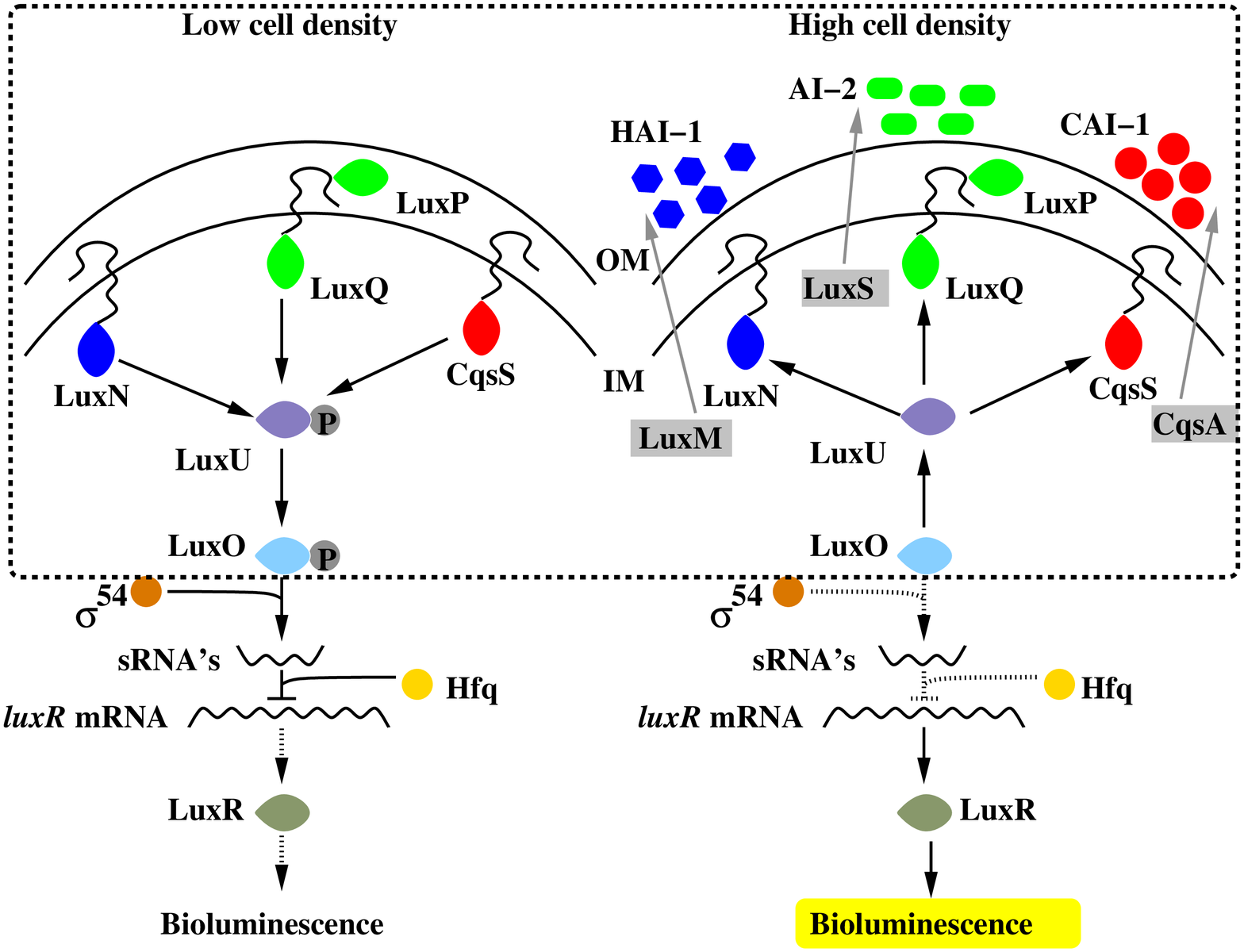}
\caption{(color online) Schematic representation of quorum sensing network in
\emph{Vibrio harveyi} at high and low cell densities. The dotted rectangle is the
input module which controls phosphorylation of LuxO in response to external
autoinducer concentrations. Solid line,
active path; Dotted line, inactive path; IM, inner membrane; OM,
outer membrane.}
\end{figure}

\section{Overview and Model}

The QS network in \emph{V. harveyi} is shown in Fig.~1.
The key upstream components of the pathway are the three sensors,
LuxN, LuxPQ and CqsS$_{Vh}$ and the corresponding autoinducer
synthases, LuxM, LuxS, and CqsA$_{Vh}$ which are responsible for
producing the three autoinducers: H-AI1, AI-2, and CAI-1,
respectively. The binding of a single autoinducer to a sensor is
highly specific, i.e., HAI-1 binds only to LuxN, AI-2 binds to
LuxPQ only, and CAI-1 binds specifically to CqsS$_{Vh}$ (see
Fig.~1). The overall network is conveniently described in terms of
functional modules.  The first (input) module includes interactions between
autoinducers ([AI$_i$] ($i=1,2,3$)) and the corresponding sensor
proteins which, through a phosphorelay mechanism, determine the
overall phosphorylation state of a $\sigma^{\mathrm 54}$-dependent response
regulator LuxO.

The second module focuses on the regulated production of sRNAs
(dependent on the phosphorylation state of LuxO) and the interaction
between the sRNAs and the master regulator protein, LuxR. The
interactions between small RNAs and their regulated targets have
been modeled in several recent studies which shed light on how target
protein expression is controlled by small RNA-mediated regulation
(Lenz et al., 2004; Levine et al., 2007; Levine and Hwa 2008; Mehta et al., 2008; Mitarai
et al., 2007). In {\em V. harveyi}, LuxR serves as the target protein
whose expression is controlled by the small RNAs in combination with the
RNA-binding protein Hfq.  The resulting
concentration of LuxR determines the level of activation or repression
of a multitude of genes including the genes involved in
bioluminescence (Waters and Bassler 2006).  The corresponding change
in the luminescence/cell output determines the luminescence profile which
is frequently used to infer network characteristics such as relative
rates of kinase/phosphotase activities by the sensor proteins (Henke
and Bassler 2004).

\begin{figure}
\label{fig2}
\includegraphics[width=0.5\linewidth,angle=0]{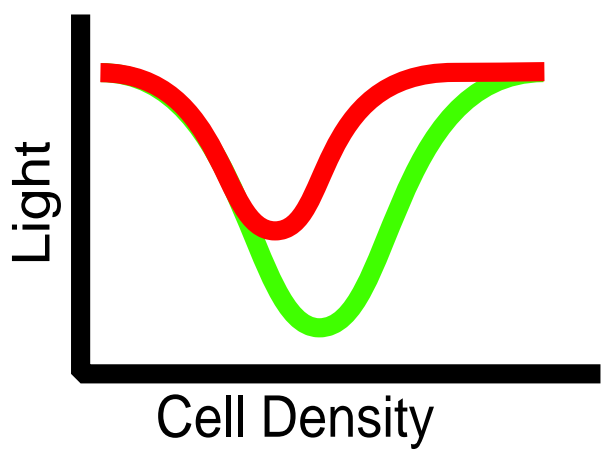}
\caption{(color online) Schematic representation of typical luminescence
curves from experiment. The green curve represents the response from a
wild type (WT) colony. The turnaround point in the curve corresponds to
cell density necessary for the activation of the genes responsible for luminescence
output per cell.
The red curve represents the luminescence/cell curve for a mutant strain that is able to achieve
the same activation at a lower cell density.}
\end{figure}

A schematic representation of typical luminescence/cell curves is shown in
Fig.~2.  Since the starting point is obtained by the dilution of cells in
the high density limit, the luminescence output per cell is maximal at
the initial time points. The luminescence output per cell then declines steadily
with increasing cell density, since luminescence genes are no longer
activated in the cells. At a specific cell density, the luminescence
curve starts to rise again signalling the start of \emph{de novo}
luminescence gene activation by cells in the growing colony. The cell
density necessary for activation can vary from the WT and mutant
strains resulting in different luminescence phenotypes (see Fig.~2).

Current data indicates that increasing cell density leads to increasing
dephosphorylation of LuxO leading to lower production rates for the
sRNAs. Correspondingly, the turnaround point in the luminescence curves
corresponds to unphosphorylated LuxO reaching a critical level above
which sRNA production is not effective at repressing LuxR levels below
the threshold for observable luminescence activation in the population of cells.
Thus, understanding how
external signals (i.e., AI concentrations as a function of cell
density) are translated into the degree of LuxO phosphorylation (i.e., the
input module) is
critical for analyzing luminescence profiles. Furthermore, pathway
mutants which function upstream of LuxO are not known to have any
direct effects on sRNA production or LuxR levels, apart from the
indirect effects mediated by LuxO. Therefore we expect that the critical
level of LuxO phosphorylation corresponding to the turnaround in
the luminescence profile is the same for all mutants. The observation
that the luminescence profiles are different for different pathway
mutants indicates different functional relations between external AI
concentrations and LuxO phosphorylation levels for the different
mutants. In the following, we derive a simple model which connects
cell density to LuxO phosphorylation and uses information from
luminescence profiles of different mutants to infer system parameters.

The sensor proteins in the QS pathway can be modeled as two state
systems (Neiditch et al., 2006; Swem et al., 2008).
We consider a further simplification which takes the sensors to be
existing either in the kinase
mode, $S_{ki}$, or in the phosphatase mode, $S_{pi}$ (where $i=1,2,3$
corresponds to the distinct sensor proteins in \emph{V. harveyi}).
In the kinase mode, the sensors can autophosphorylate and then transfer
the phosphate group to the downstream protein LuxU, whereas in the
phosphatase mode the phosphate flow is reversed.
Experiments indicate that at low cell density
(corresponding to low autoinducer concentrations) the sensors are
primarily in the kinase mode, whereas at high cell density (corresponding
to high autoinducer concentrations), the sensors are primarily in the
phosphatase mode.
Correspondingly, we consider a simplified model wherein the free sensor
corresponds to the kinase mode, whereas binding of autoinducer
results in a transition to the phosphatase mode.

At a given cell density, the external autoinducer concentrations will
be proportional to the colony forming units $N$. Since the time scale
for changes in $N$ (i.e., the doubling time) is large compared to the
time scales for binding/unbinding of ligands and subsequent
phophorylation/dephosphorylation, the corresponding reactions can be
considered in steady state for a given $N$.  Furthermore, since the
typical number of sensor proteins of each type is large, the concentration of
sensors of type $i$ is well approximated by the mean value $[S_{i}] =
c_{i} [S_{0}]$ (where $[S_{0}]$ is some reference concentration).
At a given cell density, external AI concentrations
determine the fraction of the receptors which exist in either the
kinase or phosphatase mode. For the simplest case of autoinducers
binding to their cognate sensors, we have the kinetic scheme:
\begin{equation}
\label{nhill}
S_{ki} + AI_i
\overset{k_{i}}{\underset{k_{-i}}{\rightleftharpoons}}
S_{pi},
\end{equation}

\noindent
from which the mean steady state concentrations of the sensors in either the
kinase or phosphatase mode can be obtained. More generally, to account for
cooperative effects in binding, we take the kinase/phosphatase fractions
to be:
\begin{equation}
\label{sensor}
[S_{ki}]=(1-g_i)c_{i}[S_0] \; {\rm and} \; [S_{pi}]=g_i c_{i}[S_0],
\end{equation}

\noindent
where
\begin{equation}
\label{auto}
[S_{ki}]+[S_{pi}]=c_{i}[S_0], g_i=a_i^n/(1+a_i^n), a_i=[AI_i]/\kappa_i.
\end{equation}

\noindent
and $\kappa_i = k_{-i}/k_i$.

\noindent
Equation (\ref{sensor}), with Hill coefficient $n=1$, corresponds to the
steady state
fractions for equation (\ref{nhill}), higher $n$ values correspond to
sharper switching from kinase to phosphatase mode which mimics
cooperative effects in binding. Finally, since the concentration of
the $i$-th autoinducer, [$AI_i$], is proportional to the colony forming
units (CFU), $N$, i.e.\ $[AI_i]=\nu_i N$; we renormalize the binding
constant $\kappa_i$ to define the scaled effective parameter
$\bar{\kappa}_i=\kappa_i/\nu_i$.

Typically in bacterial signal transduction, the sensor proteins in the
kinase/phosphatase modes serve as enzymes which transfer the phosphate
group to/from a response regulator protein or a phosphorelay protein
(Appleby 1996; Hoch 2000; Stock et al., 2000; Laub and Goulian 2007).
In \emph{V.\ harveyi}, this step involves phosphotransfer to the phosphorelay
protein LuxU ($U$). Phosphorylated LuxU ($U_P$) can then transfer the phosphate group
to the response regulator LuxO ($O$); similarly, unphosphorylated LuxU
serves as a receiver for removing the phosphate group from phosphorylated
LuxO ($O_P$) . We represent these processes by the following equations:
\numparts
\begin{eqnarray}
S_{ki} + U  & \overset{k_{ki}}{\rightarrow} &  S_{ki} + U_P, \\
S_{pi} + U_P &  \overset{k_{pi}}{\rightarrow} &  S_{pi} + U, \\
U_P + O & \overset{k_f}{\underset{k_b}{\rightleftharpoons}} & U + O_P.
\end{eqnarray}
\endnumparts


For the above kinetic equations, it is convenient to define key
dimensionless parameters of the model as follows
\begin{equation}
\label{dimlc}
\alpha_{ri} = c_{i} k_{ki}/k_{kr},
~\beta_i = (k_f/k_b) ( k_{ki}/k_{pi} ).
\end{equation}

\noindent
The parameter
$\alpha_{ri}$ is a measure of the relative kinase strength of $i$-th
sensor with respect to the $r$-th sensor (scaled by the mean concentrations of
the two sensors), e.g., $\alpha_{12}$ is the relative kinase strength
of sensor 2 with respect to sensor 1.  Another set of key parameters
is the ratio of the scaled kinase to phosphatase rates, $\beta_i$, of
the $i$-th sensor.
Using these dimensionless parameters, we then solve the rate equations
(4a-4c) at steady state to derive the following expression for the
fraction of unphosphorylated LuxO at steady state, $f_{\rm LuxO}=
[O]/[O]_0$ (with $[O]_0$ being the total LuxO concentration)
\begin{equation}
\label{fraco}
f_{\rm LuxO} =
\frac{\sum_i \alpha_{ri} (g_i/\beta_i)}{\sum_i \alpha_{ri} (1-g_i)
+ \sum_i \alpha_{ri} (g_i/\beta_i)} .
\end{equation}


\section{Connection to experimental data}

We now connect the model for LuxO phosphorylation developed in the
previous section to experimental luminescence curves. Recall that
the typical luminescence profile shows a well defined switching
point which signals observable {\it de novo} production of
luminescence by the population of cells. As argued earlier, this corresponds to a critical value
for the concentration of unphosphorylated LuxO. Let us denote this
critical fraction of unphosphorylated LuxO by $f^c$ and the
corresponding value of the colony forming units by $N^{c}$.
At $f_{\rm LuxO}=f^c$, for the WT luminescence curve
we have the following relation (\ref{fraco}):
\begin{equation}
\label{flfc}
\sum_i \alpha_{ri} (1-g_i) = \left ( \frac{1-f^c}{f^c} \right )
\sum_i \alpha_{ri} (g_i/\beta_i) ,
\end{equation}

\noindent
where the factors $g_{i}$ are evaluated at $N=N^{c}$. Since $N^{c}$ is known
from experiments corresponding to the WT luminescence curve, the above
equation can be regarded as a constraint on the dimensionless parameters.

We now consider the corresponding equations for luminescence
phenotypes of the mutant strains. Current knowledge of the
QS network in \emph{V. harveyi} indicates that pathway proteins
functioning upstream of LuxO primarily control LuxO phosphorylation
levels and have no direct interactions with the {\it qrr} sRNAs or the
master regulator LuxR. This suggests that for each mutant the degree of
LuxO phosphorylation needed to activate luminescence is the same
(i.e., $f^c$ is the same) since upstream proteins affect LuxR only via
LuxO-P levels. The observation that the luminescence
profiles are distinct for different pathway mutants is a consequence
of the altered functional relationship between LuxO phosphorylation
levels and external autoinducer concentrations for the mutants. Given
the defined roles of the pathway proteins, these altered functional
relationships can readily be derived within our model for all the
mutants.  For example, equation (\ref{flfc}) for the single sensor
mutant $cqsS_{Vh}$ (i.e. the strain with a deletion for the gene $cqsS_{Vh}$)
takes the form:
\begin{eqnarray*}
(1-g_1) + \alpha_{12} (1-g_2) = \left ( \frac{1-f^c}{f^c} \right )
\left [ \frac{g_1}{\beta_1} + \alpha_{12}  \frac{g_2}{\beta_2} \right ].
\end{eqnarray*}

\noindent
Note that the quantity $(1-f^c)/f^c$ can be absorbed into
the scaled kinase to phosphatase ratios $\beta_1$ and $\beta_2$. This
is equivalent to setting $f^c=1/2$ in the above equation, and since
$f^c$ is the same for all pathway mutants, a similar rescaling can be
done for the functional relationships for all the mutants. The
corresponding equations are presented in the Appendix. In the following,
we show how these equations can be used along with WT and mutant
luminescence phenotypes to determine effective system parameters and
to make testable predictions.


From the work of Henke and Bassler (2004), the critical
threshold in colony forming units ($N^c$) can be estimated for a range
of pathway mutants. The different mutant strains studied were i)
$luxN$, ii) $luxQ$, iii) $cqsS_{Vh}$, iv)
$luxN \ luxQ$, v) $luxN \ cqsS_{Vh}$, and vi)
$luxQ \ cqsS_{Vh}$. To connect the sensors of
\emph{V. harveyi} with our model, we designate sensors
LuxN, LuxQ, and CqsS$_{Vh}$ as $1$, $2$, and $3$, respectively. The
ordering of the CFU/volume for the different strains at their critical threshold
shows the following hierarchy (Henke and Bassler 2004):
\begin{equation}
\label{expord} N_{12}^c \ll N_2^c \sim N_{23}^c < N_{wt}^c < N_3^c < N_1^c \sim
N_{13}^c,
\end{equation}

\noindent
where $N_{12}^c$ is the number of colony forming units for
mutant strain $luxN \ luxQ$ at which $f_{\rm LuxO} = f^c$ and so on.
Although the values
$N_2^c$, $N_{23}^c$ and $N_1^c$, $N_{13}^c$ appear to be indistinguishable based on
available experimental data, based on the model developed we expect a small difference
in the threshold values.
For example, the difference between the {\it luxN} strain and {\it luxN cqsS$_{Vh}$} strain
is that CqsS$_{Vh}$ is active as phosphatase in the {\it luxN} mutant (close to the
switching threshold).  This implies that the switching in the luminescence phenotype
should occur at a lower $N^c$ value for the {\it luxN cqsS$_{Vh}$} strain i.e.,
$N_{1}^{c} < N_{13}^{c}$. Since CqsS$_{Vh}$ has weak effect on the
luminescence phenotype,
the switching values are indistinguishable experimentally. However to develop a consistent
model, we have to impose a small difference between the switching values
based on the constraint $N_{1}^{c} < N_{13}^{c}$ (and similarly for
$N_{2}^{c}$ and $N_{23}^{c}$).

Based on the above reasoning, we initially considered a $\sim$10\% difference between
$N_2^c$, $N_{23}^c$ and  $N_1^c$, $N_{13}^c$.
Accordingly, the values for critical thresholds (switching values,
in the units of CFU/volume) used as initial inputs were
\begin{eqnarray*}
&& N_{12}^c \sim 1 \times 10^5, N_2^c \sim 14 \times 10^5, N_{23}^c \sim 15 \times 10^5,
N_{wt}^c \sim 40 \times 10^5, \nonumber \\
&& N_3^c \sim 70 \times 10^5, N_{13}^c \sim 110 \times 10^5, N_1^c \sim 100 \times 10^5.
\end{eqnarray*}

From the discussion of the previous section, we have seen that the
input module provides us eight key parameters: two relative kinase
strengths ($\alpha_{12}$ and $\alpha_{13}$), three scaled kinase to
phosphatase ratios ($\beta_1$, $\beta_2$, and $\beta_3$) and three
effective binding constants ($\bar{\kappa}_1$, $\bar{\kappa}_2$, and
$\bar{\kappa}_3$). Given that we have experimental data for threshold
cell densities for seven strains, this indicates that if one of the
parameters is fixed, the other parameters can potentially be determined by solving
the corresponding threshold equations (see Appendix). Since previous work
indicated that the effect of CqsS$_{Vh}$ on luminescence phenotypes is minimal,
we initially fixed the parameter $\alpha_{13}$ (the relative
kinase strength of sensor 3 (CqsS$_{Vh}$) with respect to sensor 1 (LuxN)) to
0.001.\footnote{This assumption will be relaxed in the subsequent analysis
as described below}.
We then proceeded to determine the effective model parameters by solving the
threshold equations using the above experimental inputs for switching cell
densities.
We also checked the stability of the solutions to the above equations
based on small perturbations to the input parameters (data not shown).
We found that the solutions
are stable with respect to perturbations that maintain the initial $\sim$10\% difference
between $N_2^c$, $N_{23}^c$ and  $N_1^c$, $N_{13}^c$. However the solutions are
sensitive to changes in the parameters controlling the small differences in $N_c$ values.
Since experiments cannot guide us in determining the precise value of these differences,
the values of $N_2^c$ and $N_1^c$ do not serve as useful inputs in determining model
parameters. Thus additional experimental data is needed to determine model parameters
as outlined below.

The luminescence data at high cell densities (hcd) for different sensor mutants from the
work of Henke and Bassler (2004) (see Fig. 4A) provides an indirect means of estimating
model parameters.
The basic experimental observations can be summarized as follows: while the WT
strain shows a bright phenotype at hcd, the $luxS$ strain has a dim phenotype and the
$luxM$ strain has low levels of luminescence and is classified as being dark.
Furthermore the $cqsS_{Vh}$
strain has a luminescence output that is intermediate between WT and $luxS$
and the $cqsA_{Vh} \ luxS$ double mutant is dark and produces significantly
less luminescence
than a $luxM$  strain. Given our definitions of model parameters,  $f_{\rm LuxO}=1/2$
corresponds to value at which observable luminescence/cell is produced. Higher values of
 $f_{\rm LuxO}$ will correspond to brighter luminescence phenotypes, whereas a dark
luminescence phenotype implies  $f_{\rm LuxO} < 1/2 $. Thus we expect that, at hcd,
we have  $f_{\rm LuxO}$ for $luxS$ mutants to be around 0.5
(given
the dim luminescence phenotype) and  $f_{\rm LuxO}$ for the $cqsS_{Vh}$
strain to be significantly greater than
the corresponding value for the $luxS$ strain but significantly lower than 1 (the value for
the WT strain). Based on these constraints, we set the  $f_{\rm LuxO}$ values for
3 synthase mutant strains at hcd as follows:
$f_{\rm LuxO}^{cqsA} = 3/4$, $f_{\rm LuxO}^{luxm} = 1/3$ and
$f_{\rm LuxO}^{luxS \ cqsA} = 1/4$.
In combination with the expression derived for $f_{\rm LuxO}$
(equation (\ref{fraco})), these equations can be used, along with luminescence switching cell density
equations, to determine model parameters (see Appendix).

\begin{figure}
\label{fig3}
\includegraphics[width=0.75\linewidth,angle=-90]{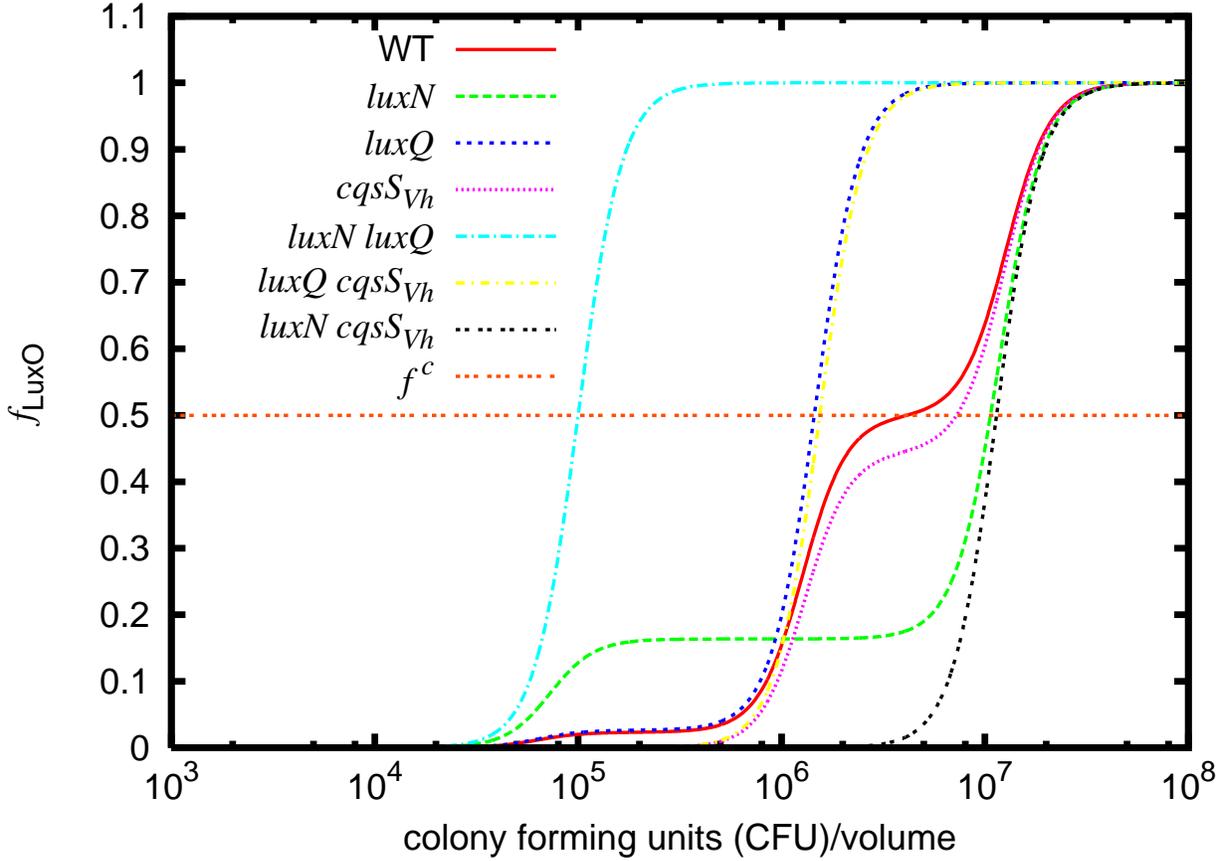}
\caption{(color online) Profile of $f_{\rm LuxO}$ as a function of colony
forming units (CFU)/volume for wild type (WT) and different sensor mutant
phenotypes. The cell density at which $f_{\rm LuxO} = f^{c}$ corresponds to
the turnaround point in the experimental luminescence curves. }
\end{figure}

First, considering equation (\ref{flfc}) for the double sensor mutants, we have the relation
between the three $\beta$-s and three $\bar{\kappa}$-s,
\begin{equation}
\label{dmut}
N_{23}^c = \bar{\kappa}_1 \beta_1^{1/n},
N_{13}^c = \bar{\kappa}_2 \beta_2^{1/n},
N_{12}^c = \bar{\kappa}_3 \beta_3^{1/n}.
\end{equation}

\noindent Also from equation (\ref{flfc}), we have the expressions for the wild
type and one single sensor mutant ($cqsS_{Vh}$) with five unknown parameters:
three kinase to phosphatase ratios ($\beta_1$, $\beta_2$ and
$\beta_3$) and two relative kinase strength ($\alpha_{12}$ and $\alpha_{13}$)
(Note that we are now considering $\alpha_{13}$ to be variable).
Using $f_{\rm LuxO}=f^c=1/2$, the switching values for WT and $cqsS_{Vh}$,
($N_{wt}^c$ and $N_3^c$) and the $f_{\rm LuxO}$ for three synthase mutants at
hcd we solve the five equations to determine the five unknown parameters.
The corresponding values for the key parameters of the model are:
$\alpha_{12} \sim 0.14$, $\alpha_{13} \sim 0.19$,
$\beta_1 \sim 8.99$, $\beta_2 \sim 0.29$ and $\beta_3 \sim 7.14$,
for the Hill coefficient $n=4$. We note that there are two sets of solutions
obtained using the above approach, however only one of these corresponds to
the experimentally observed hierarchy of switching cell densities (\ref{expord}).
Furthermore no solutions were obtained for $ n \leqslant 2$. For $n=3$, the equations
can be solved and yield parameters that are close to the those inferred for $n=4$. However
the $n=4$ results are more consistent with the experimental observation that the switching
cell densities are experimentally indistinguishable for $N_1^c$ and $N_{13}^c$ (similarly
for $N_2^c$ and $N_{23}^c$).
The high value of $n=4$
suggests that there might be cooperative effects in the switch from the
kinase to phosphatase mode for the sensors. Now using these values for
the effective parameters, we calculate the values of $f_{\rm LuxO}$ as
a function of CFU/volume (see Fig.\ 3) for the WT and different sensor mutant
phenotypes of \emph{V.\ harveyi}.
Since the effective parameters are determined, we can now use
our model to generate similar curves and make predictions for mutants that have not yet been
studied experimentally. We have checked the stability of the obtained solutions with respect to
small changes in the input values (see Appendix). We have also considered larger changes in the
input $f_{\rm LuxO}$ values consistent with the constraints noted earlier. While the precise
values of the effective
model parameters do change as the inputs are varied, there are several robust predictions
that can be made. These are discussed further in the concluding section.


\section{Conclusion and outlook}

The preceding analysis helps determine the parameters in our
minimal model. While these parameters cannot directly be compared to
experiments, they can lead to several predictions which are testable
experimentally. In the following, we outline some of the key predictions
based on our analysis.

1) The parameter $\beta_i$ is a measure of the relative kinase to
phosphatase rates for the $i$-th sensor.
Based on the values determined, the following ordering is predicted for
the relative kinase to phosphatase rates of the three sensors
LuxN $>$  CqsS$_{Vh}$  $>$LuxQ. LuxN is predicted to be the strongest kinase
which is consistent with results from previous experiments showing that LuxN
has a greater effect on LuxO phosphorylation than LuxQ (Freeman et al., 2000).
Furthermore, it is interesting
to note that recent experiments have demonstrated high kinase to
phosphatase rates for the sensor LuxN (Timmen et al., 2006).
While the corresponding value estimated by our model
($\beta_1 \sim 9 $)
cannot directly be compared to experiments since it involves
additional parameters, the ratio $\beta_i/\beta_j$ $(i \neq j)$ should
correspond to experimental estimation of the ratio of kinase to
phosphatase rates of two sensors.  From our model
we consistently find that $\beta_2/ \beta_1 \ll 1$
and $\beta_2/ \beta_3 \ll 1$Ê indicating the the effective kinase to phosphatase
activity ratio for LuxQ is much lower than the other two sensors.
Note that this prediction differs significantly from the previous characterization
(Henke and Bassler 2004)
that kinase to phosphatase activity ratio for LuxQ is greater than that of CqsS$_{Vh}$.
It would thus be of interest to carry out experiments to measure relative kinase to
phosphatase rates for the sensors LuxQ and CqsS$_{Vh}$ to see if the predictions
are borne out.

2) Experiments with mutant strains (besides those used as inputs to
our model) indicate that at high cell densities, the luminescence
phenotypes can be broadly categorized into 3 types: dark, dim and bright.
Since $f^c = 1/2$ is the threshold for luminescence activation in our
model, we take these categories to correspond to the following: dark
($0 \leqslant f_{\rm LuxO} < 0.4$), dim ($0.4 < f_{\rm LuxO} < 0.6$) and
bright ($0.6 < f_{\rm LuxO} \leqslant 1.0$). Using these criteria, we can
now predict the luminescence phenotypes at high cell density for other
pathway mutants (i.e.\ those not included in the experimental inputs used
to determine model parameters). The corresponding results are listed in
Table~\ref{predict}. We note that all mutant
strains with LuxM deleted ($luxM$) are dark. This is consistent
with previous experimental results (Freeman and Bassler 1999).
Other interesting predictions are \\
i) While $cqsA_{Vh} \ luxN$ is bright (comparable to $cqsA_{Vh}$) at hcd,
the strain $cqsA_{Vh} \ luxQ$ is predicted to be dark; \\
ii) $luxS$ is brighter than $luxM$ at hcd , however $cqsA_{Vh} \ luxS$
is predicted to be darker than $cqsA_{Vh} \ luxM$ (note that
this is  consistent with the observations in Henke and Bassler (2004)).

\begin{table}
\caption{\label{predict} Predictions for luminescence output per cell of
different synthase mutants and mixed sensor-synthase mutants. }
\begin{indented}
\item[] \begin{tabular}{@{}ll}
\br
Phenotype & Mutant \\
\mr
dark &  $luxM$, $luxM \ luxS$, $luxS \ cqsA_{Vh}$, $luxM \ cqsA_{Vh}$, \\
 & $luxN \ luxS$, $luxQ \ luxM$, $luxQ \ cqsA_{Vh}$, $cqsS_{Vh} \ luxM$ \\
 \mr
dim &  $luxS$, $cqsS_{Vh} \ luxS$ \\
\mr
bright & $cqsA_{Vh}$, $luxN \ cqsA_{Vh}$
 \\
\br
\end{tabular}
\end{indented}
\end{table}

It should be noted that the results presented in Fig.~3 are just for sensor
mutants whereas Table~\ref{predict} is for synthase mutants and mixed sensor-synthase
mutants. For the different mutants given in Table~\ref{predict}, the maximal value of the
$f_{\rm LuxO}$ curve differs from 1 and stays within the defined range
(according to the broad categories discussed in the paper) even at the hcd
in contrast to the behavior shown in Fig.~3 for the sensor mutants.

3) To figure out the values of the effective parameters of the model, we have
used the switching value ($N^c$) of WT, $cqsS_{Vh}$ and double
sensor mutants
from the experiment (Henke and Bassler 2004). With these derived values
of the effective parameters, we can now  predict the switching values of the other
two bright sensor mutant strains ($luxN$ and $luxQ$) at hcd
(in the units of CFU/volume),
\begin{eqnarray*}
N_1^c \sim 100 \times 10^5, N_2^c \sim 14 \times 10^5.
\end{eqnarray*}

\noindent
It is interesting to note that the above switching values are in good agreement
with the observation that $N_1^c$ is experimentally indistinguishable from $N_{13}^c$
and $N_2^c$ is experimentally indistinguishable from $N_{23}^c$
(see Fig. 3).
In addition, the effective parameter set predicts the switching values
($N^c$, in units of CFU/volume)
for the two bright mutant strains $cqsA_{Vh}$ and $luxN \ cqsA_{Vh}$
mentioned in Table~\ref{predict} as $\sim 130 \times 10^5$ and
$\sim 156 \times 10^5$, respectively.

4) Recent experiments have probed the response of the QS
pathway to externally controlled autoinducer concentrations
(Mok et al., 2003). In these experiments, the autoinducer
production is switched off by deleting the corresponding synthases
and then autoinducers are added back exogenously in controlled
amounts.  In our model this behavior can be mimicked
by controlling the quantity $g_i$ in equation (\ref{auto}). For each
synthase mutation the autoinducer production is switched off so that
$g_i = 0$ as $AI_i = 0$ ($i=1,2,3$). As autoinducers are added to
the network from outside, the quantity $g_i$ grows and tends to one
as $AI_i \rightarrow \infty$.
For this setup, our analysis indicates a situation wherein the sensor
CqsS$_{Vh}$ plays an important role in regulating the response which is
contrary to what is normally assumed. Consider the situation for which {\emph all} the
autoinducer synthases have been deleted and subsequently saturating
amounts of  $AI_1$ are added. In this case, we predict a significant
difference between the luminescence output per cell for the two cases corresponding
to i) low external $AI_3$ concentrations and (ii) high external
$AI_3$ concentrations. The difference between these two cases is that
the sensor CqsS$_{Vh}$ is primarily in kinase mode for case (i) and in
phosphatase mode for case (ii). Our analysis thus suggests a testable
prediction for an experimentally realizable situation wherein signaling
through CqsS$_{Vh}$ significantly changes the output from the QS pathway.


5) Finally, we examine predictions from our model for the expression of genes
that are also controlled by $f_{\rm LuxO}$ through LuxR but
are not directly related to luminescence/cell. Waters and Bassler
(2006) studied several genes regulated by LuxR and classified them into
different categories based on the activation/repression induced by the
presence of high concentrations of either AI$_1$ or AI$_2$ or both.
We will focus on the category of genes (labeled ``class 3" genes)
which are defined as genes that show an equally notable change in
expression when either AI$_1$ and/or AI$_2$ are present in high concentrations.
Within our model, we can
calculate the the values of $f_{\rm LuxO}$ for the 3 cases : (i) High
concentration of AI$_1$ only, (ii) high concentration of AI$_2$ only and (iii)
high concentration of both AI$_1$ and AI$_2$. Out of these the lowest value of
$f_{\rm LuxO}$ corresponds to case (ii) i.e., high concentration of AI$_2$
only. Since class 3 genes are fully activated/repressed when high
concentrations of AI$_2$ only are present, it follows that the $f^c$
for all genes in this category must be lesser than the value of $f_{\rm LuxO}$
when only AI$_2$ levels are high ($f_{\rm LuxO} = 0.33$). (Note that we have assumed that
AI$_3$ levels are at high concentrations in the above experiments since they
are at high cell densities).  This observation indicates that
an upper bound for activation/repression of class 3 genes corresponds to
$f^c=0.33$. Using this, the following testable predictions can be made

\begin{itemize}
    \item{The synthase mutant $luxM$ can fully activate/repress
          class 3 genes at high cell density.
         Note that luminescence genes, in
        contrast, are not activated at high cell density in a $luxM$
        mutant. }
    \item{Similarly, the sensor-synthase mutants
         $luxM \ cqsS_{Vh}$
         and $luxQ \ cqsA_{Vh}$ cannot activate luminescence genes at
         high cell density whereas they are predicted to fully
          activate/repress all class 3 genes at high cell density
        }.
\end{itemize}

The minimal model presented in this work can be generalized further as
more experimental data becomes available.  An important generalization
would be to relax some of the assumptions made by considering a two-state
model (Swem et al., 2008) which incorporates non-zero phosphatase
activity in the \emph{on} (free) state and nonzero kinase
activity in the \emph{off} (bound) state. We note that this will add several
additional parameters to our current model. With additional experimental data,
the generalized model could be used to estimate the expanded set of
effective parameters. While the effective parameters so determined are likely
to be different from the values determined using the minimal model, the
framework connecting the model parameters to experimental data will
essentially be the same.

In summary, we have proposed a minimal model to study the quorum
sensing network in \emph{V. harveyi}.  Using experimental data for
luminescence phenotypes of WT and different mutant strains, we provide
a framework to estimate the effective dimensionless parameters of the
model.  Correspondingly, the model can be used to predict the
luminescence phenotypes of other pathway mutants which have not been
experimentally studied to date. The proposed framework captures the
key features of the signal transduction in \emph{V.\ harveyi} and can
contribute to guiding and interpreting experimental efforts analyzing
the QS pathway in the Vibrios.


\ack SKB and RVK acknowledge support from the institutes ICTAS and
IBPHS at Virignia Tech and the ASPIRES award (Virginia Tech). ATF
acknowledges support from NSF IGERT grant DGE-0504196.


\section*{Appendix}

\appendix

\setcounter{section}{1}

For the relative kinase strength ($\alpha_{ri} = c_i k_{ki}/k_{kr}$ for
$i=1,2,3$) of the sensors, we generally use the kinase strength of LuxN,
i.e., $k_{k1} (r=1)$, as the reference kinase. Now using equation (\ref{flfc})
we explicitly write the functional relation for WT strain evaluated at
$N=N^c_{wt}$ for $f_{\rm LuxO} = f^c$:
\begin{equation}
\fl
(1-g_1) + \alpha_{12} (1-g_2) + \alpha_{13} (1-g_3)
= \left ( \frac{1-f^c}{f^c} \right )
\left [ \frac{g_1}{\beta_1} + \alpha_{12} \frac{g_2}{\beta_2} +
\alpha_{13} \frac{g_3}{\beta_3}  \right ].
\label{eq:WT_switch}
\end{equation}

\noindent Similarly for $luxN$ mutants we use kinase strength
of LuxQ, i.e., $k_{k2} (r=2)$, as the reference kinase whereas for
$luxQ$ and $cqsS_{Vh}$ we use kinase strength
of LuxN as the reference kinase as in WT. Thus the functional relations
for the single sensor mutants $luxN$, $luxQ$ and $cqsS_{Vh}$
evaluated at $N^c_1$, $N^c_2$ and $N^c_3$, respectively, are:

\noindent For $luxN$ (r=2):
\begin{equation}
(1-g_2) + \frac{\alpha_{13}}{\alpha_{12}} (1-g_3)
= \left ( \frac{1-f^c}{f^c} \right )
\left [  \frac{g_2}{\beta_2} + \frac{\alpha_{13}}{\alpha_{12}} \frac{g_3}{\beta_3}  \right ] .
\end{equation}

\noindent For $luxQ$ (r=1):
\begin{equation}
(1-g_1) + \alpha_{13} (1-g_3)
= \left ( \frac{1-f^c}{f^c} \right )
\left [  \frac{g_1}{\beta_1} + \alpha_{13} \frac{g_3}{\beta_3}  \right ] .
\end{equation}

\noindent For $cqsS_{Vh}$ (r=1):
\begin{equation}
(1-g_1) + \alpha_{12} (1-g_2)
= \left ( \frac{1-f^c}{f^c} \right )
\left [  \frac{g_1}{\beta_1} + \alpha_{12} \frac{g_2}{\beta_2}  \right ] .
\label{eq:cqsS_switch}
\end{equation}

\noindent For double sensor mutants, the value of the relative kinase strengths
becomes 1 as there is only 1 sensor. Hence the functional relations
for the double sensor mutants $luxN \ luxQ$,
$luxQ \ cqsS_{Vh}$ and $luxN \ cqsS_{Vh}$
evaluated at $N^c_{12}$, $N^c_{23}$ and $N^c_{13}$, respectively, are:

\noindent For $luxN \ luxQ$ (r=3):
\begin{equation}
(1-g_3) = \left ( \frac{1-f^c}{f^c} \right ) \frac{g_3}{\beta_3} .
\end{equation}

\noindent For $luxQ \ cqsS_{Vh}$ (r=1):
\begin{equation}
(1-g_1) = \left ( \frac{1-f^c}{f^c} \right ) \frac{g_1}{\beta_1} .
\end{equation}

\noindent For $luxN \ cqsS_{Vh}$ (r=2):
\begin{equation}
(1-g_2) = \left ( \frac{1-f^c}{f^c} \right ) \frac{g_2}{\beta_2} .
\end{equation}


\noindent Now, using equation (\ref{dmut}) and the expression for the functional relation for the WT
strain given above (A.1) we have the following equation evaluated at $N=N_{wt}^c$ for $f^c=1/2$,
\begin{eqnarray}
\label{eq:WT_switch-nc}
&& \left [ 1 - \left ( \frac{N_{wt}^c}{N_{23}^c} \right )^n \right ]
+ \alpha_{12} \frac{1+(N_{wt}^c/\bar{\kappa}_1)^n}{1+(N_{wt}^c/\bar{\kappa}_2)^n}
\left [ 1 - \left ( \frac{N_{wt}^c}{N_{13}^c} \right )^n \right ] \nonumber \\
&& + \alpha_{13} \frac{1+(N_{wt}^c/\bar{\kappa}_1)^n}{1+(N_{wt}^c/\bar{\kappa}_3)^n}
\left [ 1 - \left ( \frac{N_{wt}^c}{N_{12}^c} \right )^n \right ] = 0.
\end{eqnarray}

\noindent Similarly, for $luxN$, $luxQ$ and $cqsS_{Vh}$ strains we have the following set of equations evaluated at $N_1^c$, $N_2^c$ and $N_3^c$, respectively, for $f^c=1/2$,
\begin{eqnarray}
&& \left [ 1 - \left ( \frac{N_1^c}{N_{13}^c} \right )^n \right ]
+ \frac{\alpha_{13}}{\alpha_{12}} \frac{1+(N_1^c/\bar{\kappa}_2)^n}{1+(N_1^c/\bar{\kappa}_3)^n}
\left [ 1 - \left ( \frac{N_1^c}{N_{12}^c} \right )^n \right ] = 0, \\
&& \left [ 1 - \left ( \frac{N_2^c}{N_{23}^c} \right )^n \right ]
+ \alpha_{13} \frac{1+(N_2^c/\bar{\kappa}_1)^n}{1+(N_2^c/\bar{\kappa}_3)^n}
\left [ 1 - \left ( \frac{N_2^c}{N_{12}^c} \right )^n \right ] = 0, \\
&& \left [ 1 - \left ( \frac{N_3^c}{N_{23}^c} \right )^n \right ]
+ \alpha_{12} \frac{1+(N_3^c/\bar{\kappa}_1)^n}{1+(N_3^c/\bar{\kappa}_2)^n}
\left [ 1 - \left ( \frac{N_3^c}{N_{13}^c} \right )^n \right ] = 0.
\label{eq:cqsS_switch-nc}
\end{eqnarray}


\noindent To find the unknown parameters of the system of equations ($\alpha_{12}$, $\alpha_{13}$, $\beta_1$, $\beta_2$, and $\beta_3$), we use equations (\ref{eq:WT_switch-nc}) and (\ref{eq:cqsS_switch-nc}) evaluated at $N=N_{wt}^c$ and $N=N_{3}^c$, respectively, along with the following three equations all evaluated at $N=N^{\rm large}$:

\begin{equation}
f_{\rm LuxO}^{luxM} = \frac{\alpha_{12}(g_2/\beta_2)+\alpha_{13}(g_3/\beta_3)
} {
1+\alpha_{12}(1-g_2)+\alpha_{13}(1-g_3)+\alpha_{12}(g_2/\beta_2)
+\alpha_{13}(g_3/\beta_3)
},
\label{eq:luxM_SS}
\end{equation}

\begin{equation}
f_{\rm LuxO}^{cqsA} = \frac{ (g_1/\beta_1) + \alpha_{12} (g_2/\beta_2)
} {
\alpha_{13}+(1-g_1)+\alpha_{12}(1-g_2)+ (g_1/\beta_1)+\alpha_{12} (g_2/\beta_2)
},
\label{eq:cqsA_SS}
\end{equation}

\begin{equation}
f_{\rm LuxO}^{luxS \ cqsA} = \frac{ (g_1/\beta_1)
} {
\alpha_{12}+\alpha_{13}+(1-g_1) + (g_1/\beta_1)
}.
\label{eq:luxScqsA_SS}
\end{equation}

\noindent
Equations (\ref{eq:luxM_SS}-\ref{eq:luxScqsA_SS}) give the $f_{\rm LuxO}$
values for the three mutants $luxM$, $cqsA_{Vh}$, and $luxS \ cqsA_{Vh}$ once the
system has reached steady-state for $N=N^{\rm large}$ such that $f_{\rm LuxO}$ has saturated. 
Equations (\ref{eq:WT_switch-nc}) and (\ref{eq:cqsS_switch-nc}-\ref{eq:luxScqsA_SS}) 
are then numerically solved using Mathematica
(Wolfram Research, Inc., Version 6, 2008) which yielded two solutions subject
to the constraint that all the parameters must be real and positive. We keep
the solution that best agrees with experimental data (see main text). When solving these
equations, we used $f_{\rm LuxO}^{luxM}=0.33$, $f_{\rm LuxO}^{cqsA}=0.75$,
$f_{\rm LuxO}^{luxS \ cqsA}=0.25$ and $n=4$.

We next analyzed the changes to the solutions based on small
perturbations to the input parameters. Each perturbation for the input values
is drawn from a random Gaussian distribution whose mean is the base value
and variance is the base value$\times \sigma$, where $\sigma$ is chosen
such that 68\% (98\%) of the perturbed values lie within 2\% (5\%) of the
base value. For example, to generate a list of perturbed $N_{12}^c$ values,
we set the mean of the Gaussian distribution to be $N_{12}^c$ and the
variance to be $N_{12}^c \times \sigma$, etc. Using this scheme, we
generated 100 random data points for the input values (the switching values)
and numerically solve equations 
(\ref{eq:WT_switch-nc}) and (\ref{eq:cqsS_switch-nc}-\ref{eq:luxScqsA_SS})
with $n=4$ to generate the effective parameters. Note, $f_{\rm LuxO}^{luxM}$,
$f_{\rm LuxO}^{cqsA}$, $f_{\rm LuxO}^{luxS \ cqsA}$ are also perturbed in the
same fashion.

\begin{figure}
\label{figa1}
\includegraphics[width=0.9\linewidth,angle=0]{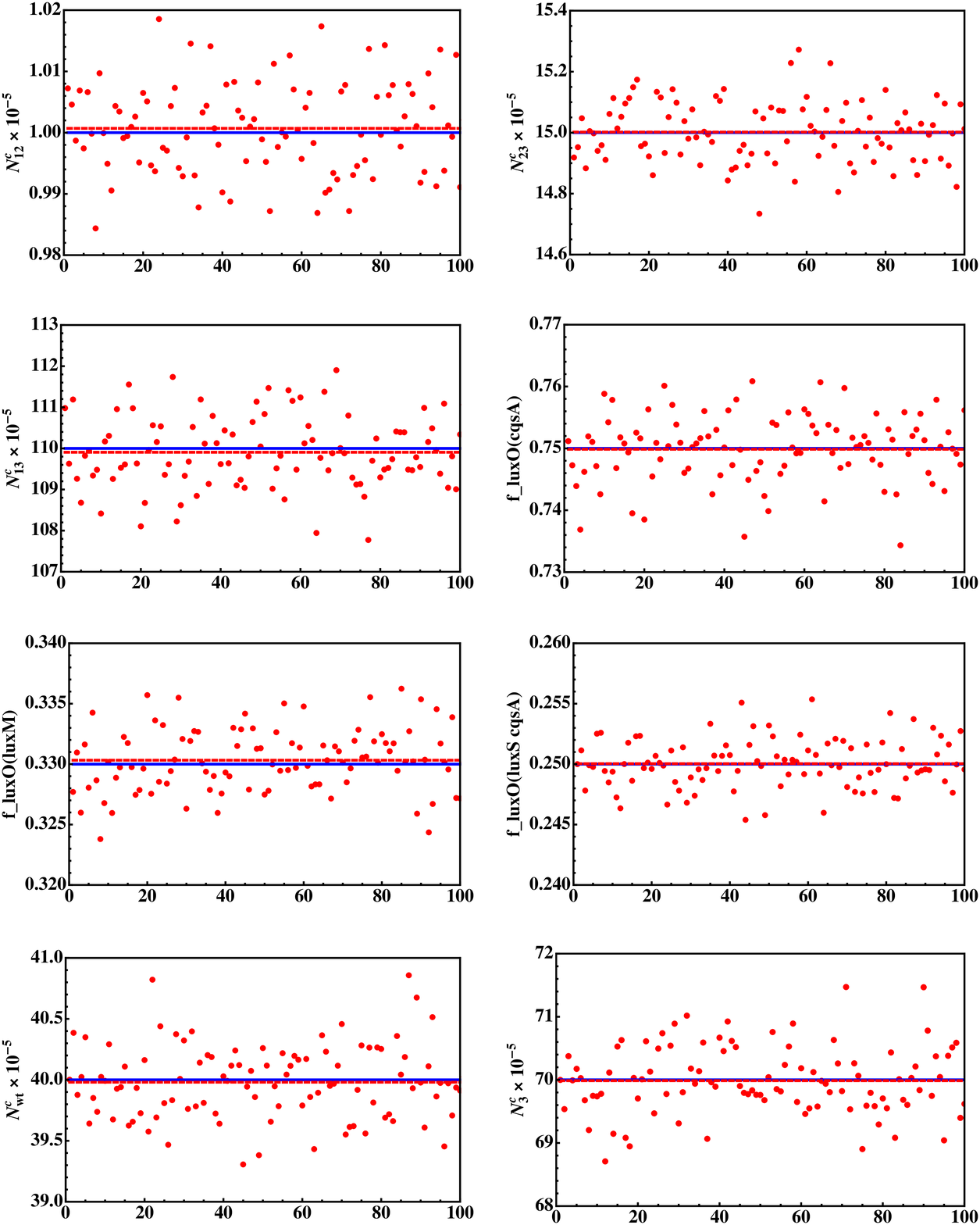}
\caption{
(color online) Results of sensitivity analysis for the input base values
The blue line represents the unperturbed data and the red dashed line is
the mean of the 100 perturbed data points represented by scattered red
points.
 }
\end{figure}

\begin{figure}
\label{figa2}
\includegraphics[width=0.9\linewidth,angle=0]{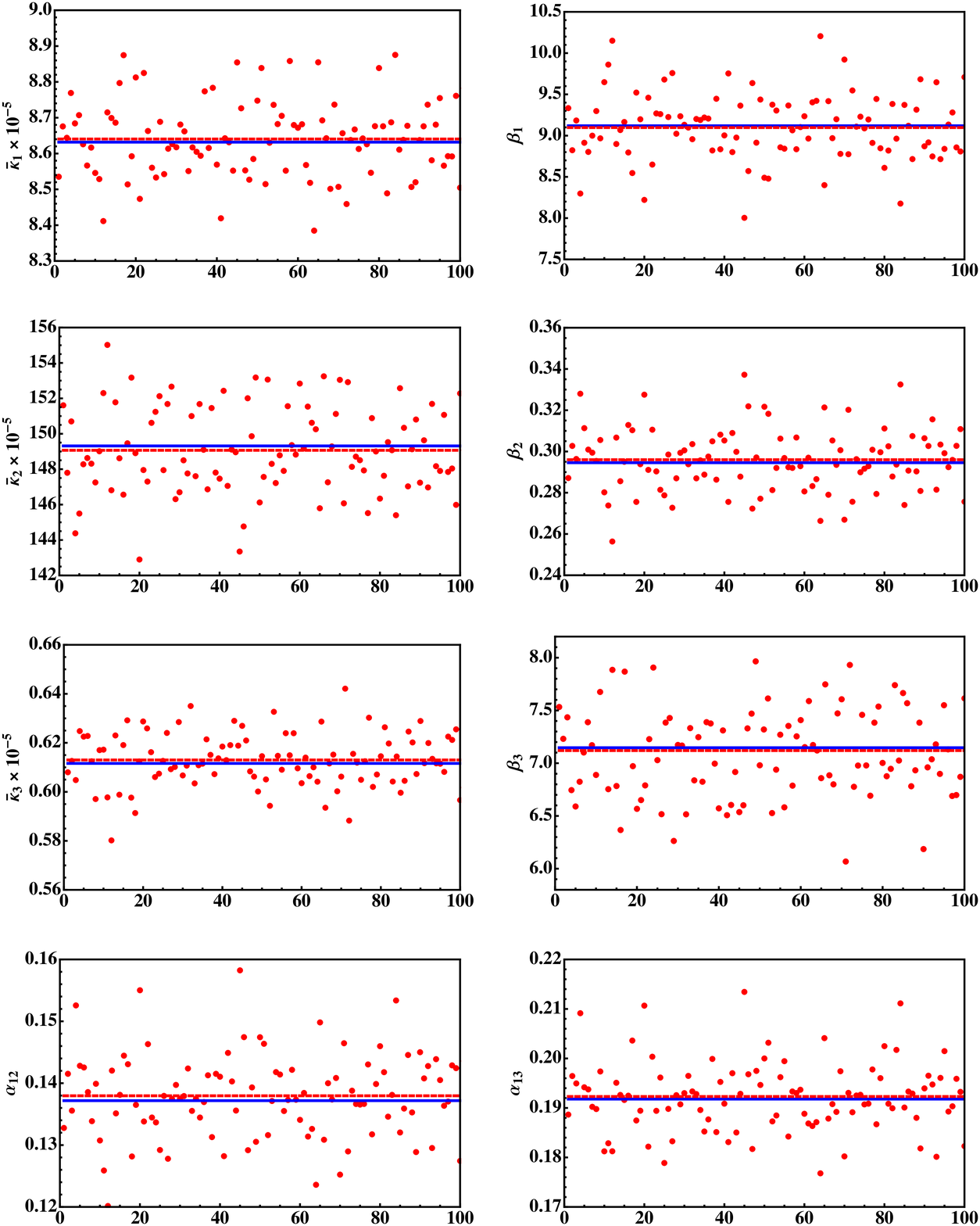}
\caption{
(color online) Results of sensitivity analysis for the effective parameters.
The blue line, red line and red scattered points have the same meaning
as in Fig.~A1.
}
\end{figure}

The resultant data of the sensitivity analysis are shown in
Figs.~A1-A2. The nature of the data shown in Figs.~A1-A2 suggests that
the parameter set obtained using the experimental switching values from
Henke and Bassler (2004) is robust against small perturbations.


\section*{Glossary}

\noindent \emph{Quorum sensing}. Process leading to regulation of gene expression in response to cell density.

\noindent \emph{Autoinducers}. Small signaling molecules produced by bacteria which bind to specific receptors and induce the quorum sensing response.

\noindent \emph{Kinase}. Enzyme acting as phosphate donor.

\noindent \emph{Phosphatase}.  Enzyme acting as phosphate acceptor.

\noindent \emph{Bioluminescence}. Production of light by living organism as a result of
internal chemical reactions.

\noindent \emph{Vibrio harveyi}. Gram-negative and bioluminescent marine bacterium.


\section*{References}

\begin{harvard}

\item[] Appleby J L, Parkinson J S and Bourret R B 1996
Signal transduction via the multi-step phosphorelay:
Not necessarily a road less traveled
\emph{Cell} \textbf{86} 845

\item[] Bassler B L and Losick R 2006
Bacterially speaking
\emph{Cell} \textbf{125}  237

\item[] Bjarnsholt T and Givskov M 2007
Quorum-sensing blockade as a strategy for enhancing host defences against bacterial pathogens
\emph{Philos. Trans. R Soc. Lond. B} \textbf{362} 1213

\item[] Freeman J A and Bassler B L 1999
A genetic analysis of the function of LuxO, a two-component response regulator involved
in quorum sensing in \emph{Vibrio harveyi}
\emph{Mol. Microbiol.} \textbf{31} 665

\item[] Freeman J A, Lilley B N and Bassler B L 2000
A genetic analysis of the functions of LuxN: a two-component hybrid sensor kinase that
regulates quorum sensing in \emph{Vibrio harveyi}
\emph{Mol. Microbiol.} \textbf{35} 139

\item[] Henke J M and Bassler B L 2004
Three parallel quorum-sensing systems regulate gene expression in \emph{Vibrio harveyi}
\emph{J. Bacteriol.} \textbf{186} 6902

\item[] Hoch J A 2000
Two-component and phosphorelay signal transduction
\emph{Curr. Opin. Microbiol.} \textbf{3} 165

\item[] Laub M T and Goulian M 2007
Specificity in two-component signal transduction pathways
\emph{Annu. Rev. Genet.} \textbf{41} 121

\item[] Lenz D H, Mok K C, Lilley B N, Kulkarni R V, Wingreen N S and Bassler B L 2004
The small RNA chaperone Hfq and multiple small RNAs control quorum sensing in
\emph{Vibrio harveyi} and \emph{Vibrio cholerae}
\emph{Cell} \textbf{118} 69

\item[] Levine E, Zhang Z, Kuhlman T and Hwa T 2007
Quantitative characteristics of gene regulation by small RNA
\emph{PLoS Biol.} \textbf{5} e229

\item[] Levine E and Hwa T 2008
Small RNAs establish gene expression thresholds
\emph{Curr. Opin. Microbiol.} \textbf{11} 574

\item[] Mehta P, Goyal S and Wingreen N S 2008
A quantitative comparison of sRNA-based and protein-based gene regulation.
\emph{Mol. Syst. Biol.} \textbf{4} 221

\item[] Miller M B and Bassler B L 2001
Quorum sensing in bacteria
\emph{Annu. Rev. Microbiol.} \textbf{55} 165

\item[] Mitarai N, Andersson A M, Krishna S, Semsey S and Sneppen K 2007
Efficient degradation and expression prioritization with small RNAs
\emph{Phys. Biol.}  \textbf{4} 164

\item[] Mok K C, Wingreen N S and Bassler B L 2003
\emph{Vibrio harveyi} quorum sensing: a coincidence detector for two autoinducers controls gene expression
\emph{EMBO J.} \textbf{22} 870

\item[] Nealson K H, Platt T and Hastings J W 1970
Cellular control of the synthesis and activity of the bacterial luminescent system
\emph{J. Bacteriol.} \textbf{104} 313

\item[] Neiditch M B, Federle M J, Pompeani A J, Kelly R C, Swem D L, Jeffrey P D, Bassler B L and Hughson F M 2006
Ligand-induced asymmetry in histidine sensor kinase complex regulates quorum sensing
\emph{Cell} \textbf{126} 1095

\item[] Stock A M, Robinson V L and Goudreau P N 2000
Two-component signal transduction
\emph{Annu. Rev. Biochem.} \textbf{69} 183

\item[] Swem L R, Swem D L, Wingreen N S and Bassler B L 2008
Deducing receptor signaling parameters from in vivo analysis: LuxN/AI-1 quorum sensing in \emph{Vibrio harveyi}
\emph{Cell} \textbf{134} 461

\item[] Timmen M, Bassler  B L and Jung K 2006
AI-1 influences the kinase activity but not the phosphatase activity of LuxN of \emph{Vibrio harveyi}
\emph{J. Biol. Chem.} \textbf{281} 24398

\item[] Tu K C and Bassler B L 2007
Multiple small RNAs act additively to integrate sensory information and control quorum sensing in \emph{Vibrio harveyi}
\emph{Genes. Dev.} \textbf{21} 221

\item[] Waters C M and Bassler B L 2005
Quorum sensing: Cell-to-cell communication in bacteria
\emph{Annu. Rev. Cell. Dev. Biol.} \textbf{21} 319

\item[] Waters C M and Bassler B L 2006
The \emph{Vibrio harveyi} quorum-sensing system uses shared regulatory components to discriminate between multiple autoinducers
\emph{Genes. Dev.} \textbf{20} 2754

\end{harvard}

\end{document}